\documentclass[aps,prl,twocolumn,showpacs,preprintnumbers,amsmath,amssymb]{revtex4}    

\usepackage{graphicx}% Include figure files
\usepackage{dcolumn}% Align table columns on decimal point
\usepackage{bm}% bold math
\usepackage{amsfonts}
\usepackage{amsmath,amssymb}%%%%%%%%% JN's definitions:

\newcommand {\beq} {\begin{equation}}
\newcommand {\eeq} {\end{equation}}
\newcommand {\beqa}{\begin{eqnarray}}
\newcommand {\eeqa}{\end{eqnarray}}

\begin{document}

%%%%%%%%%%%%%%%%%%%%%%%%%%%%%%%%%%%%%%%%%%%%%%%%%%%%%%%%%%%%%%%%%%%%
%  TITLE / AUTHOR                                                  %
%%%%%%%%%%%%%%%%%%%%%%%%%%%%%%%%%%%%%%%%%%%%%%%%%%%%%%%%%%%%%%%%%%%%

\title{
Monte Carlo studies
of Matrix theory correlation functions
}
 
\author{Masanori Hanada$^{1}$}
\email{masanori.hanada@weizmann.ac.il}
%\cite{EmailKN}
\author{Jun Nishimura$^{2,3}$}
\email{jnishi@post.kek.jp}
\author{Yasuhiro Sekino$^{4}$}
\email{ysekino@v101.vaio.ne.jp}
\author{Tamiaki Yoneya$^{5}$}
\email{tam@hep1.c.u-tokyo.ac.jp}

%\cite{EmailJN}

%\address{
\affiliation{
$^{1}$Department of Particle Physics and Astrophysics, 
Weizmann Institute of Science, Rehovot 76100, Israel\\
$^{2}$KEK Theory Center, High Energy Accelerator Research Organization, 
		Tsukuba 305-0801, Japan \\
$^{3}$Department of Particle and Nuclear Physics,
School of High Energy Accelerator Science,
%\\
Graduate University for Advanced Studies (SOKENDAI),
Tsukuba 305-0801, Japan\\
%${}^3$Department of Particle and Nuclear Physics, 
%		The Graduate University for Advanced Studies (SOKENDAI),
%			Tsukuba, Ibaraki 305-0801, Japan
$^{4}$Okayama Institute for Quantum Physics, 
1-9-1 Kyoyama, Okayama 700-0015, Japan\\
$^{5}$Institute of Physics, University of Tokyo, 
Komaba, Meguro-ku, Tokyo 153-8902, Japan
}

\date{November, 2009; preprint: WS/14/09/NOV-DPPA, KEK-TH-1337, OIQP-09-12, UT-Komaba/09-5
%, hep-th/yymmnnn
%\today %%new
}% It is always \today, today,

%%%%%%%%%%%%%%%%%%%%%%%%%%%%%%%%%%%%%%%%%%%%%%%%%%%%%%%%%%%%%%%%%%%%
%  ABSTRACT    	                                                   %
%%%%%%%%%%%%%%%%%%%%%%%%%%%%%%%%%%%%%%%%%%%%%%%%%%%%%%%%%%%%%%%%%%%%

\begin{abstract}
We study 
correlation functions in 
$(0+1)$-dimensional maximally supersymmetric U($N$) gauge theory,
which represents the low-energy effective theory of D0-branes. 
In the large-$N$ limit, the gauge-gravity duality
predicts 
power-law behaviors in the infrared region for the
two-point correlation functions of operators corresponding to 
supergravity modes. 
We evaluate such correlation functions on the gauge theory side 
by the Monte Carlo method.
Clear power-law behaviors are observed at $N=3$, 
and the predicted exponents are confirmed consistently. Our results suggest that the agreement extends to the M-theory regime, where the supergravity analysis in 10 dimensions 
may not be justified \emph{a priori}.
\end{abstract}

%11.25.-w Theory of fundamental strings
%11.25.Sq Nonperturbative techniques; string field theory

\maketitle

%%%%%%%%%%%%%%%%%%%%%%%%%%%%%%%%%%%%%%%%%%%%%%%%%%%%%%%%%%%%%%%%%%%%
%  1. INTRODUCTION                                                 %
%%%%%%%%%%%%%%%%%%%%%%%%%%%%%%%%%%%%%%%%%%%%%%%%%%%%%%%%%%%%%%%%%%%%

\paragraph*{Introduction.---}

Maximally supersymmetric Yang-Mills theories (SYM) 
in various dimensions 
are important because of their connection to non-perturbative
formulations of string theory. 
Of particular interest is the (0+1)D SYM with $U(N)$ gauge symmetry,
which is supposed to describe the low-energy gravitational
dynamics of D0-branes in 10D type IIA superstring theory.
In an appropriate large-$N$ limit, this theory
was proposed to be a definition of M-theory in a special 
light-like frame, and it is commonly 
referred to as the Matrix theory~\cite{BFSS}. 
M-theory is a hypothetical 11D theory \cite{mtheory},
whose low energy effective theory is given by 11D supergravity,
and it is believed to appear in the strong coupling limit of 
10D type IIA superstring theory.

Indeed it was confirmed that 
scattering amplitudes in the Matrix theory 
at weak coupling are consistent 
with predictions from 11D supergravity.
Even the three-body force, which is 
a characteristic non-linear effect of general relativity,
has been reproduced~\cite{okawayoneya} 
from the quantum loop effects of massive open strings connecting D0-branes. 
On the other hand, direct perturbative calculations of correlation functions,
which could provide crucial information on the as yet mysterious theory,
are plagued by severe infrared divergences
caused by the massless modes
inherent in the theory.
We therefore need genuinely nonperturbative methods to study 
such quantities.

In the case of
${\cal N}$=4 SYM in (3+1) dimensions, 
various useful insights have been gained from 
the AdS/CFT correspondence 
\cite{Maldacena:1997re, GKPW}.
This conjectural duality enables us to study the ${\cal N}=4$ SYM 
in the large-$N$ limit with large 't Hooft coupling constant
by using the weakly coupled supergravity, which describes 
the low-energy limit of the string theory.
Likewise, SYM in ($p$+1)-dimensions,
corresponding to the world-volume theory of D$p$-branes,
is expected to be
dual to a superstring theory
on the D$p$-brane background
in the near-horizon limit \cite{IMSY, Jevicki}.

Assuming this duality for the $p=0$ case and 
using the known dictionary~\cite{Taylor} 
between the 
supergravity modes and the Matrix theory operators,
one may hope to calculate the correlation functions 
in the large-$N$ limit, 
on the basis of the Gubser-Klebanov-Polyakov-Witten prescription~\cite{GKPW}. 
This program was carried out a decade ago 
by Y.S.\ and T.Y.~\cite{SY}
using the ``generalized conformal symmetry'' \cite{Jevicki, Y} 
as a guide for classifying and interpreting the obtained results. 
The prediction is that the two-point correlation functions obey
the power law
\begin{eqnarray}
\Big\langle
{\cal O}(t){\cal O}(t')
\Big\rangle
\propto
\frac{1}{|t-t'|^{2\nu+1}} \ ,
\label{coordspace}
\end{eqnarray}
for a set of operators ${\cal O}(t)$ 
corresponding to the supergravity modes. 
The exponents $\nu$ are fractional numbers 
which differ from the canonical values, and they are 
related to the generalized conformal dimensions $\Delta$ 
of the operators under consideration 
by $\Delta=-1+\frac{10}{7}\nu$ \cite{SY}. 
The power-law behavior 
 conforms to the existence of a unique 
threshold bound state at zero energy \cite{wittenbound} 
in the gauge theory, 
which corresponds to single-body states in the graviton 
supermultiplet.

This result has been re-derived in later works \cite{ASY}, 
extending the analysis in Ref.\ \cite{SY} 
to general D$p$-branes ($p<5$) in 
the PP-wave limit. There it has also been 
shown that the infrared behavior of the correlation functions 
for operators corresponding to excited stringy modes 
is typically $\exp\{-c\big(g_{{\rm YM}}^2N|t-t'|^{3}\big)^{1/5}\}$ 
up to power corrections. 

The near-horizon limit of the D0-brane background is valid when 
the radial coordinate $r$ in the bulk supergravity satisfies
$r\ll (g_sN)^{1/7}$ in the string unit $\alpha'=1$. The analysis using the supergravity 
approximation is then possible when 
$g_sN \gg 1$, and is trustable when both the 
background curvature and the effective string coupling 
given by the dilaton expectation value
are small. 
The string coupling becomes stronger towards the center, 
while the curvature becomes larger towards the boundary.
Taking into account the fact that the radial coordinate $r$ 
is related to the time scale $t$  at the boundary 
as $ r \sim (t/\lambda^{1/2})^{-2/5}$~\cite{SY} with 
$\lambda\equiv g_{{\rm YM}}^2N \sim g_sN$, 
the conditions for justifying 
the supergravity result (\ref{coordspace}) 
can be summarized as \cite{IMSY, Jevicki, SY}
\begin{equation}
\lambda^{-1/3} \ll |t-t'|\ll
\lambda^{-1/3} N^{10/21} \ .
\label{region} 
\end{equation} 
Hence, the application of the 
gauge-gravity correspondence 
is legitimate when one studies the region 
which is sufficiently infrared compared to the length scale
set by $\lambda$. Since 
the upper bound in \eqref{region} diverges at large $N$, 
the infrared behavior of the correlation 
functions 
can be predicted 
reliably from the supergravity analysis 
in the large-$N$ limit. 

In this Letter we evaluate the correlation functions on the gauge theory side 
by the Monte Carlo method \cite{Hanada-Nishimura-Takeuchi} 
and compare the results with the 
predictions from the gauge-gravity correspondence. 
Related Monte Carlo analyses have been recently applied 
by the authors including M.H.\ and J.N.\
to reproduce the black hole thermodynamics \cite{AHNT,endnote}
and the Schwarzschild radius \cite{HMNT}
of the dual geometry.  
To the best of our knowledge,
this is the first attempt to compute the Matrix theory correlation functions 
from first principles~\cite{endnote2}.
Our Monte Carlo data are consistent with 
the supergravity predictions with reasonable accuracy
despite 
the fact that the matrix size is as small as $N=3$.
Possible interpretations shall be given later.

%%%%%%%%%%%%%%%%%%%%%%%%%%%%%%%%%%%%%%%%%%%%%%%%%%%%%%%%%%%%%%%%%%%%
%  2. SIMULATION TECHNIQUES                                        %
%%%%%%%%%%%%%%%%%%%%%%%%%%%%%%%%%%%%%%%%%%%%%%%%%%%%%%%%%%%%%%%%%%%%

\paragraph*{Simulating Matrix theory.---}

The action of the Matrix theory 
can be written (in the Euclidean convention) as
\begin{eqnarray}
S &=&
\frac{N}{\lambda}\int_0^\beta dt\ {\rm Tr}
\Bigl\{
\frac{1}{2}(D_tX_i)^2
-
\frac{1}{4}[X_i,X_j]^2
\nonumber\\
& &
+
\frac{1}{2}\psi_\alpha D_t\psi_\alpha
-
\frac{1}{2}\psi_\alpha(\gamma_i)_{\alpha\beta}
[X_i,\psi_\beta]
\Bigl\} \ , 
\end{eqnarray} 
where $X_i\ (i=1,\cdots , 9)$ and $\psi_\alpha\ (\alpha=1,\cdots,16)$ are 
$N\times N$ bosonic and fermionic Hermitian matrices, on which  
the covariant derivative $D_t$ acts as 
$D_t=\partial_t-i[A,\ \cdot\ ]$ with the gauge field $A$.
The $16 \times 16$ matrices $\gamma_i$ satisfy the Clifford algebra 
$\{ \gamma_i,\gamma_j \}= 2\delta_{ij}$.
It is convenient to adopt units in which $\lambda=1$
without loss of generality. This does 
not contradict the conditions for the bulk theory, 
since the gauge theory has no independent length 
scale other than the 
coupling constant $g_{{\rm YM}}^{-2/3}$ and hence
the strong coupling 
limit amounts to the IR limit. 

In order to put the system on a computer,
we have to introduce the UV and IR cutoffs appropriately.
The extent $\beta$ in the Euclidean time direction 
represents the IR cutoff, which should be sufficiently
large in order to see the correct infrared properties.
Since we are not interested in the finite temperature behaviors 
unlike in Refs. \cite{AHNT,HMNT},
we impose periodic boundary conditions for both 
bosonic and fermionic matrices, respecting supersymmetry. 
In order to introduce the UV cutoff,
we first fix the gauge as
$A(t)=\frac{1}{\beta}{\rm diag}(\alpha_1,\cdots,\alpha_N)$,
where $-\pi<\alpha_i\le \pi$, and then
introduce a Fourier mode cutoff $\Lambda$ as
$X_i(t)=\sum_{n=-\Lambda}^\Lambda\tilde{X}_{in}e^{i\omega nt}$ and 
$\psi_\alpha(t)=
\sum_{n=-\Lambda}^\Lambda\tilde{\psi}_{\alpha n}e^{i\omega nt}$, 
where $\omega=2\pi/\beta$. 
Integration over the fermionic matrices 
yields a Pfaffian, 
which could be complex in general.
In this work, we simply neglect its phase 
following the argument in Ref.\ \cite{HMNT}. 
While we do not have purely theoretical justification of our assumption
at this moment,
the previous works \cite{AHNT,HMNT} provide empirical evidence 
that the phase does not 
affect the results in the continuum limit. 
Including its effect is possible in principle
by the so-called reweighting method, but it is not feasible for our 
purpose, since the demand on the statistics would increase exponentially
with the system size due to huge cancellation. This is generally called the ``sign problem''. 
The system of finite degrees of freedom obtained under the above 
assumption can be simulated by using the
Rational Hybrid Monte Carlo algorithm \cite{RHMC} 
as described in Ref.\ \cite{AHNT}. 
%%%%%%%%%%%%%%%%%%%%%%%%%%%%%%%%%%%%%%%%%%%%%%%%%%%%%%%%%%%%%%%%%%%%
%  3. SIMULATION RESULTS                                           %
%%%%%%%%%%%%%%%%%%%%%%%%%%%%%%%%%%%%%%%%%%%%%%%%%%%%%%%%%%%%%%%%%%%%

\paragraph*{Correlation functions.---}

First let us study a series of 
operators $J^{+ij}_{l,i_1,\cdots,i_l}$ ($l\ge 1$),
which is defined by 
\begin{eqnarray}
J^{+ij}_{l,i_1,\cdots,i_l}
\equiv
\frac{1}{N} {\rm Str}
\left(
F_{ij}X_{i_1}\cdots X_{i_l}
\right) \ , 
\label{Jplus}
\end{eqnarray}
where $F_{ij}\equiv -i[X_i,X_j]$ and 
${\rm Str}$ represents the symmetrized trace treating $F_{ij}$ 
as a single unit.
The value of $\nu$ in (\ref{coordspace}) predicted in this case 
is $\nu=2l/5$ \cite{SY}.

\begin{figure}[htb]
\begin{center}
\rotatebox{-90}{
\includegraphics[height=6cm]{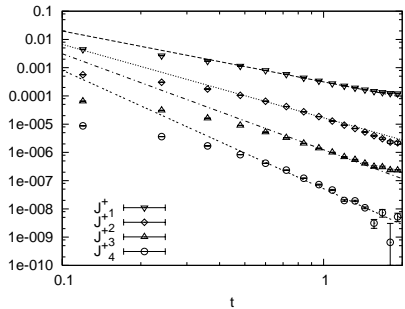}
}
\end{center}
\caption{
The real-space two-point correlation functions 
$\langle J_l^+(t)J_l^+(0)\rangle$ $(l=1,2,3,4)$ 
are plotted for $N=3$ and $\beta=4$. The UV cutoff is $\Lambda=16$. 
The straight lines are fits to the predicted power-law behavior.
}
\label{J^+}
\end{figure}

Since we are dealing with the Fourier modes in our simulation,
the two-point functions which are directly accessible
are those in the momentum space
$\left\langle 
\tilde{{\cal O}}(p)
\tilde{{\cal O}}(-p)
\right\rangle$,
where $p= n \, \omega$.
Correlation functions in the real space can be 
obtained by the inverse Fourier transformation
$
\left\langle
{\cal O}(t)
{\cal O}(0)
\right\rangle
=
\frac{1}{\beta}
\sum_p
\left\langle
\tilde{{\cal O}}(p)
\tilde{{\cal O}}(-p)
\right\rangle
e^{ipt} $. 
However, the results obtained in this way
oscillates with the frequency of O($\Lambda \, \omega $)
as a result of the Gibbs phenomenon.
For the series of operators (\ref{Jplus}),
we find that the momentum-space correlator 
falls off as  
$p^{-2}$
at large $p$. 
Assuming that this behavior continues to infinite $p$,
we extend it up to $p=1000 \, \omega $,
and then make the inverse
Fourier transformation, which removes
the Gibbs phenomenon completely \cite{Hanada-Nishimura-Takeuchi}.
In Fig.~\ref{J^+}, we show the real-space correlation functions 
obtained in this way.
The predicted power-law behavior can be seen 
in the region $0.5\lesssim t\lesssim 1.5$,
which is somewhat wider than the criterion \eqref{region} 
for justifying the supergravity analysis. 
Note that the IR bound 
of \eqref{region} is $|t-t'| \ll 1.69$ for $N=3$. 
Considering that 1.5 is close to $\beta/2=2$, we expect that
the deviation observed at $t \gtrsim 1.5$ is rather due to the IR cutoff.

%%%%% J^+_i series. %%%%%
%%%%%%%%%%%%%%%%%%%

\begin{figure}[htb]
\begin{center}
\rotatebox{-90}{
\includegraphics[height=6cm]{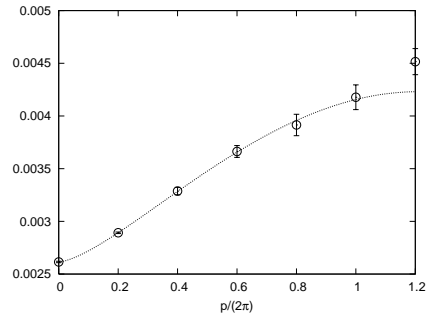}
}
\end{center}
\caption{
The momentum-space two-point correlation function 
$\left\langle\tilde{T}^+_2(p)\tilde{T}^+_2(-p)
\right\rangle$ 
is plotted for $N=3$ and $\beta=5$. 
The dotted line represents a fit 
to the behavior (\ref{asmp-p})
treating $\nu$ as a free parameter 
in the range $0\le p/(2\pi)\le 1$. 
The best fit is obtained for $\nu = 0.80 \pm 0.03$. 
}
\label{T^+_2}
\end{figure}

Let us next study the operator $T^{+i}_{2,jk}$ defined by 
\begin{eqnarray}
T^{+i}_{2,jk}\equiv
\frac{1}{N} {\rm Str}
\left(
(D_t X_i)X_j X_k
\right) \ , 
\label{Tplus}
\end{eqnarray}
for which $\nu$ is predicted to be $\nu=4/5$ \cite{SY}. 
Unlike the previous case, 
the two-point function in the Fourier space does not fall off as 
$p^{-2}$ at large $p$. 
This is likely due to the presence of a derivative $D_t$ in (\ref{Tplus}).
Furthermore, we observe considerable dependence on the UV cutoff $\Lambda$.
We therefore evaluate the two-point function 
in the momentum space at various $\Lambda$
and make an extrapolation to $\Lambda=\infty$ at each $p$. 
The leading finite $\Lambda$ effects are consistent with $1/\Lambda$ 
as one might 
expect theoretically, and we make an extrapolation 
with $\Lambda=6,8,12$ for $0\le p/(2\pi)\le 0.6$ and 
$\Lambda=8,12,16$ for $p/(2\pi) \ge 0.8$.
Figure \ref{T^+_2} shows the results obtained in this way
for $N=3$, $\beta=5$.

On the other hand, 
the prediction (\ref{coordspace}) in the Fourier space reads \cite{SY} 
\begin{eqnarray}
\left\langle
\tilde{{\cal O}}(p)
\tilde{{\cal O}}(-p)
\right\rangle
\simeq f(p) +  |p|^{2\nu} g(p) 
\label{T^+_2 mom space}
\end{eqnarray}
at small $p$,
where $f(p)$ and $g(p)$ are analytic functions ($g(0)\ne 0$), 
which cannot be determined by the supergravity analysis.
The odd powers of $p$ in $f(p)$ and $g(p)$ are 
forbidden by the 
time reflection invariance.
%parity invariance.
For the present operator, considering that $\nu=4/5$,
a few leading terms 
at small $p$ 
are predicted to be of the form
\begin{eqnarray}
\left\langle\tilde{T}^+_2(p)\tilde{T}^+_2(-p)\right\rangle
\simeq
a+b p^2 + c |p|^{2\nu} \ , 
\label{asmp-p}
\end{eqnarray}
where $a$, $b$ and $c$ are constants. Note that the polynomial 
terms 
do not affect the long-distance behavior 
of the correlation function. 
Treating $\nu$ as a free parameter and fitting our data
in $0\le p/(2\pi) \le 1$ to 
(\ref{asmp-p}), 
we obtain $\nu = 0.80 \pm 0.03$ as 
in Fig.\ \ref{T^+_2}. 

%%%%%%%%%%%%%%%%%%%

\begin{figure}[htb]
\begin{center}
\rotatebox{-90}{ 
\includegraphics[height=6cm]{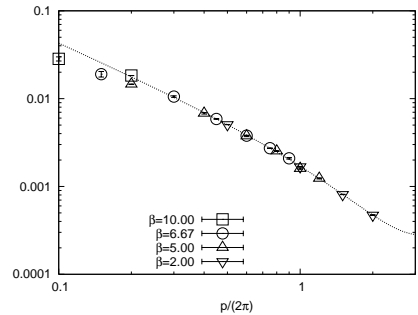}}
\end{center}
\caption{
The momentum-space two-point correlation function 
$\left\langle\tilde{T}^{++}_2(p)
\tilde{T}^{++}_2(-p)
\right\rangle$  
is plotted in the log-log scale for $N=3$. 
The dotted line represents a fit 
to the behavior (\ref{asmp-p2})  
treating $\nu$ as a free parameter
in the range $0.3 \le p/(2\pi) \le 2$.
The best fit is obtained for $\nu = -0.61 \pm 0.02$. 
}
\label{T^++_2}
\end{figure}

As the last example,  let us consider the operator 
\beq
T^{++}_{2,ij}\equiv\frac{1}{N} {\rm tr} (X_i X_j) \quad (i\neq j) \ ,
\eeq
for which the supergravity analysis predicts $\nu=-3/5$ \cite{SY}. 
The two-point function is predicted to behave as
$| t - t'|^{1/5}$ in the real space,
which is divergent as $|t-t'|\rightarrow \infty$.
This IR divergence is reminiscent of the free-field behavior 
$|t-t'|^2$ albeit with much weaker power. 
Here we examine the behavior (\ref{T^+_2 mom space}) for small $p$
in the Fourier space.
Considering that $\nu=-3/5$,
a few leading terms are predicted to be of the form 
\begin{eqnarray}
\left\langle\tilde{T}^{++}_2(p)\tilde{T}^{++}_2(-p)\right\rangle
\simeq
a |p|^{2\nu} + b + c |p|^{2\nu+2} \ . 
\label{asmp-p2}
\end{eqnarray}
Note that the leading term is divergent as $p \rightarrow 0$, as opposed to 
the previous cases with vanishing behaviors.
In Fig.~\ref{T^++_2}, we show a log-log plot of the two-point function.
We extrapolated our data to $\Lambda=\infty$ as we did for $T^+_2$. 
Treating $\nu$ as a free parameter and fitting our results 
to (\ref{asmp-p2})
in the region $0.3 \le p/(2\pi) \le 2$, we obtain 
$\nu = -0.61 \pm 0.02$. 
The deviation from the behavior (\ref{asmp-p2})
at small $p$
is most likely due to
finite IR cutoff effects as one can see from the trends
with increasing $\beta$.

%%%%%%%%%%%%%%%%%%%

%%%%%%%%%%%%%%%%%%%%%%%%%%%%%%%%%%%%%%%%%%%%%%%%%%%%%%%%%%%%%%%%%%%%
%  Conclusion and Discussions                                      %
%%%%%%%%%%%%%%%%%%%%%%%%%%%%%%%%%%%%%%%%%%%%%%%%%%%%%%%%%%%%%%%%%%%%
\paragraph*{Summary and discussion.---}
 
In this Letter we presented the first Monte Carlo evaluation of
two-point correlation functions of the supergravity operators in the Matrix theory.
In particular, 
we observed that the power law predicted by the gauge-gravity correspondence
continues to hold in the infrared region 
beyond the naive criterion for the validity of the 
supergravity analysis. 

From the gauge theory point of view, 
the existence of a threshold bound state at zero energy 
requires the power law at sufficiently large $|t-t'|$ 
for any $N\, (\ge 2)$. 
In fact, the existence of a unique threshold bound state 
for each $N \, (\ge 2)$ has been  
argued, 
based on the calculation of the Witten index \cite{threbound}.
Let us also consider the spectral representation of the two-point
function
\beq
\left\langle 
\tilde{{\cal O}}(p)
\tilde{{\cal O}}(-p)
\right\rangle = \int d \mu  \, \rho(\mu) \, \frac{1}{p^2+\mu^2} \ .
\eeq
The power law of the two-point function at small $p$
implies that the spectral density $\rho(\mu)$
behaves as $\rho(\mu)\sim \mu^{2\nu+1}$ at small $\mu$. 
The fact that the exponent $\nu$ increases in general \cite{SY}
with the angular momentum 
carried by the operator
is consistent with the continuous mass spectrum 
of the intermediate many-body states composed of the zero-energy bound states. 

On the gravity side, let us emphasize that the upper bound in \eqref{region} 
is 
merely
a sufficient condition for justifying the 
supergravity analysis.
Here we recall that the wave function of the supergravity fluctuations 
obtained in Ref.\ \cite{SY} 
is exponentially small
in the central region. 
Therefore, 
it is possible that the 10D supergravity approximation in this region 
is not affected by the increasingly 
large effective string coupling at small $r$ and hence at large $|t-t'|$. 
On the other hand, 
in the short-distance regime $|t-t'| \ll \lambda^{-1/3}$, 
where the curvature becomes large on the gravity side, 
we expect large 
deviations from the power-law behavior \eqref{coordspace}.
This is clear, in particular, for the cases with $2\nu+1>0$
since there should be no UV singularity in one dimension. Such deviations are indeed observed
in the stringy PP-wave analysis  \cite{ASY} on the bulk side. 

Another significant aspect
of our results is that the exponents obtained from the 
Monte Carlo data at $N=3$ essentially coincide with
the predictions in the large-$N$ limit.
We have also studied the $N=2$ case, but the exponents
are unaltered within numerical uncertainties.
This suggests that the exponents are 
independent of $N$ without  O$(1/N^2)$ corrections, 
possibly due to the BPS nature of 
the supergravity modes. 
Furthermore, since the exponents 
cannot depend on the coupling constant $\lambda$,
the same exponents should be valid in the M-theory regime, 
where we consider the large-$N$ limit with fixed $g_s \sim g^2_{{\rm YM}}$
and 
with the scaling 
$|t-t'| \sim  N|\tau-\tau'|$ 
of the light-cone time (in the string unit), corresponding to the 
infinite momentum limit $P^+ =N/g_s \rightarrow \infty$. 
For possible interpretations of the predicted exponents 
from the 11 dimensional viewpoint of 
the Matrix theory conjecture,
we refer the reader to Ref.\ \cite{Y} in addition to \cite{SY}.
More details of our analysis as well as results on other correlation functions
shall be reported in a separate paper. 

%%%%%%%%%%%%%%%%%%%%%%%%%%%%%%%%%%%%%%%%%%%%%%%%%%%%%%%%%%%%%%%%%%%%
%  ACKNOWLEDGEMENTS                                                %
%%%%%%%%%%%%%%%%%%%%%%%%%%%%%%%%%%%%%%%%%%%%%%%%%%%%%%%%%%%%%%%%%%%%

\paragraph*{Acknowledgments.---}

We thank M.~Terashi for his contribution
at the early stage of this work.
We are also grateful to O.~Aharony 
and Y.~Kikukawa for useful discussions 
and comments. 
Computations have been carried out on PC clusters 
at KEK 
and Yukawa Institute.
The present work is supported in part by Grant-in-Aid 
for Scientific Research 
(No.\ 19340066 and 20540286 for J.N.,
No.\ 21740216 for Y.S.
and No.\ 20340048 for T.Y.) 
from Japan Society for the Promotion of Science.

%%%%%%%%%%%%%%%%%%%%%%%%%%%%%%%%%%%%%%%%%%%%%%%%%%%%%%%%%%%%%%%%%%%%
%  REFERENCE                                                      %
%%%%%%%%%%%%%%%%%%%%%%%%%%%%%%%%%%%%%%%%%%%%%%%%%%%%%%%%%%%%%%%%%%%%

%\end{references}


\begin{thebibliography}{99}

%\begin{references}
%\vspace*{-1.5cm}
%\bibitem[\ast]{EmailKN} Electronic mail: kawahara@post.kek.jp 
%\bibitem[\star]{EmailJN} Electronic mail: jnishi@post.kek.jp 

\bibitem{BFSS}
T.~Banks, W.~Fischler, S.~H.~Shenker and L.~Susskind,
Phys.\ Rev.\ D {\bf 55}, 5112 (1997).
For a comprehensive review of the Matrix theory, 
see, {\it e.g.}, W.~Taylor, Rev. Mod. Phys. {\bf 73}, 419 (2001). 

\bibitem{mtheory}
E.~Witten,
Nucl.\ Phys.\ B {\bf 443}, 85 (1995).

\bibitem{okawayoneya} Y.~Okawa and T.~Yoneya, 
Nucl. Phys.\ {\bf B538}, 67 (1999); {\bf B541}, 163 (1999).

\bibitem{Maldacena:1997re}
J.~M.~Maldacena, 
Adv.\ Theor.\ Math.\ Phys.\  {\bf 2}, 231 (1998).


\bibitem{GKPW}
S.~S.~Gubser, I.~R.~Klebanov and A.~M.~Polyakov,
Phys.\ Lett.\  B {\bf 428}, 105 (1998);
E.~Witten,
Adv.\ Theor.\ Math.\ Phys.\  {\bf 2}, 253 (1998).

%% \bibitem{AdSreview}
%%   O.~Aharony, S.~S.~Gubser, J.~M.~Maldacena, H.~Ooguri and Y.~Oz,
%%   Phys.\ Rept.\  {\bf 323}, 183 (2000).


\bibitem{IMSY}
N.~Itzhaki, J.~M.~Maldacena, J.~Sonnenschein and S.~Yankielowicz,
Phys.\ Rev.\ D {\bf 58}, 046004 (1998).

\bibitem{Jevicki}
A.~Jevicki and T.~Yoneya,
Nucl.\ Phys.\  B {\bf 535}, 335 (1998).

\bibitem{Taylor}
D.~N.~Kabat and W.~Taylor,
Phys.\ Lett.\  B {\bf 426}, 297 (1998);
W.~Taylor and M.~Van Raamsdonk,
JHEP {\bf 9904}, 013 (1999);
%  W.~Taylor and M.~Van Raamsdonk,
Nucl.\ Phys.\  B {\bf 558}, 63 (1999).

\bibitem{SY}
Y.~Sekino and T.~Yoneya,
Nucl.\ Phys.\  B {\bf 570}, 174 (2000); 
Y.~Sekino,
Nucl.\ Phys.\  B {\bf 602}, 147 (2001)


\bibitem{Y}
T. ~Yoneya, Classi.\  Quantum Gravity {\bf 17}, 1307(2000).   

\bibitem{wittenbound} E.~Witten, Nucl. Phys.\ {\bf 460}, 335 (1996). 

\bibitem{ASY}
M.~Asano, Y.~Sekino and T.~Yoneya,
Nucl.\ Phys.\  B {\bf 678}, 197 (2004);
M.~Asano and Y.~Sekino,
Nucl.\ Phys.\  B {\bf 705}, 33 (2005);
M.~Asano,
JHEP {\bf 0412}, 029 (2004).

%% \bibitem{endnote}
%% In fact, analysis in the plane wave limit shows that for operators 
%% which correspond to excited string states, correlator goes like
%% $\exp\{-c(g_{YM}^2N)^{-1/5}|t-t'|^{3/5}\}$~\cite{ASY}.


\bibitem{Hanada-Nishimura-Takeuchi}
M.~Hanada, J.~Nishimura and S.~Takeuchi,
Phys.\ Rev.\ Lett.\  {\bf 99}, 161602 (2007).

\bibitem{AHNT}
K.~N.~Anagnostopoulos,
M.~Hanada, J.~Nishimura and S.~Takeuchi,
Phys.\ Rev.\ Lett.\  {\bf 100}, 021601  (2008);
M.~Hanada, Y.~Hyakutake, J.~Nishimura and S.~Takeuchi,
Phys.\ Rev.\ Lett.\  {\bf 102}, 191602 (2009). 

\bibitem{endnote}
%See also,
See also Ref.\ \cite{CW} for Monte Carlo simulations
based on the lattice regularization.
%
%
\bibitem{CW}
%\cite{Catterall:2007fp}
%\bibitem{Catterall:2007fp}
  S.~Catterall and T.~Wiseman,
  %``Towards lattice simulation of the gauge theory duals to black holes and hot
  %strings,''
  JHEP {\bf 0712}, 104 (2007);
%  [arXiv:0706.3518 [hep-lat]].
  %%CITATION = JHEPA,0712,104;%%
%
%\cite{Catterall:2008yz}
%\bibitem{Catterall:2008yz}
%  S.~Catterall and T.~Wiseman,
  %``Black hole thermodynamics from simulations of lattice Yang-Mills theory,''
  Phys.\ Rev.\  D {\bf 78}, 041502 (2008);
%  [arXiv:0803.4273 [hep-th]].
  %%CITATION = PHRVA,D78,041502;%%
%
%\cite{Catterall:2009xn}
%\bibitem{Catterall:2009xn}
%  S.~Catterall and T.~Wiseman,
  %``Extracting black hole physics from the lattice,''
  arXiv:0909.4947 [hep-th].
  %%CITATION = ARXIV:0909.4947;%%


\bibitem{HMNT}
M.~Hanada, A.~Miwa, J.~Nishimura and S.~Takeuchi,
Phys.\ Rev.\ Lett.\  {\bf 102}, 181602 (2009).

\bibitem{endnote2}
See Ref.\ \cite{Hiller} for 
a numerical analysis based
on the discrete light-cone quantization
in the $(1+1)$-dimensional case.
%, numerical analysis based
%on the discrete light-cone quantization was done in~

\bibitem{Hiller}
J.~R.~Hiller, S.~S.~Pinsky, N.~Salwen and U.~Trittmann,
Phys.\ Lett.\  B {\bf 624}, 105 (2005).

\bibitem{RHMC}
%\cite{Clark:2003na}
%\bibitem{Clark:2003na}
M.~A.~Clark and A.~D.~Kennedy,
  %``The RHMC algorithm for 2 flavors of dynamical staggered fermions,''
  Nucl.\ Phys.\ Proc.\ Suppl.\  {\bf 129}, 850 (2004).
%  [arXiv:hep-lat/0309084].
  %%CITATION = NUPHZ,129,850;%%
%% M.~A.~Clark, A.~D.~Kennedy and Z.~Sroczynski,
%% Nucl.\ Phys.\ Proc.\ Suppl.\  {\bf 140}, 835 (2005).
  
\bibitem{threbound} P.~Yi, Nucl.\ Phys.\ {\bf B505}, 307 (1997); 
S.~Sethi and M.~Stern, Comm.\ Math.\ Phys.\ {\bf 194}, 675 (1998); 
G.~Moore, N.~Nekrasov and S.~Shatashvili, 
{\it ibid} {\bf 209}, 77 (2000) and references therein. 

\end{thebibliography}
\end{document}